\documentstyle[12pt]{article}
\begin{document}
\input epsf
\begin{center}
THE LORENTZ-DIRAC EQUATION AND THE STRUCTURES OF SPACETIME\\
\bigskip
Manoelito Martins de Souza \footnote{
Universidade Federal do Esp\'{\i}rito Santo\\
Departamento de F\'{\i}sica \\
 29060-900  Vit\'oria - ES - Brasil \\
e-mail: manoelit@cce.ufes.br}\\
\end{center}
\medskip
\begin{center}
ABSTRACT
\end{center}
A new interpretation of the causality implementation in the Lienard-Wiechert
solution raises new doubts against the validity of the Lorentz-Dirac equation
and  the limits of validity of the Minkowski structure of spacetime.

\section{INTRODUCTION}
Finding the correct equation of motion for a pointlike charged classical
particle was, early in this century, a major problem in theoretical  physics.
The proposed third-order Lorentz-Dirac equation could not be accepted because
of its numerous problems. These problems have not been solved but just
forgotten since with the advent of quantum mechanics came also the hope that
they could be properly understood in the scope of a quantum theory . This
represented, in the point of view of this paper, a bad corner-stone for
theoretical physics: for not solving them, one has failed on seeing that the
Minkowski space is not the appropriate underlying geometric  structure for the
description of close interacting fields.
The solution to these problems is still of great relevance since it may signal
steering corrections one has to make in field theory for avoiding old problems
of QED and the stalling situations found today in, for example, quantum gravity
and QCD.\\
  In modern field theories Poincar\'e invariance is imposed, and  then
the Min{\-}kowski space-time is taken as the appropriate scenario  for
describing non-gravitational phenomena. For electromagnetic fields in vacuum,
far from charges, this has received confirmation from  a solid experimental
basis , but not for fields in  a  close vicinity  of their sources. Even from a
theoretical viewpoint the question  is  not  so  clear: the problems quantum
field theories face for dealing  with  fields
defined in  close  neighboring  points  are  well  known.  These
difficulties are generally taken as indications of some failures in
the quantum basis of the theories or of an scale on its limit of validity. In
this  letter we want first to emphasize that this is the  same  problem that
occurs  in  classical  physics disguised on this  old controversy  about  the
correct  equation  of motion for the classical electron. Having inherited the
same spacetime structure of their classical predecessors, it is not a surprise
that  the quantum theories also face a similar problem for defining fields in a
too close vicinity. Therefore,  the roots of this problem must be searched at
deeper grounds,  in  the very foundation of the assumed structures of the
space-time continuum.

A classical spinless point charge in an  isotropic  and  homogenous
Poincar\'e invariant space-time and the validity of energy  momentum
conservation$^{(1,2,3)}$ lead  unequivocally to the Lorentz-Dirac equation,
\begin{equation}
\label{lordireq}
ma = eF_{ext} .V +\frac{2e^{2}}{3}(\stackrel{.}{a}-a^{2}V).
\end{equation}
This equation is written in a context where the electron world-line, parameterized  by
its proper time $\tau$, is a known function,$z(\tau).$ Then, the electron velocity is $V=dz/d\tau$ $a=dV/d\tau$, and ${\dot a}\equiv da/d\tau$. $eF_{ext}$ is the exterior force driving the electron, which, if taken as of electromagnetic origin, is put as $F^{\mu}_{ext}=F^{\mu\nu}_{ext}V_{\nu}.$ m and e are the electron mass and charge, respectively. c is the speed of light.
The presence of the Schott term, $\frac{2e^{2}}{3}\!\stackrel{.}{a}$, is  the
cause  of  some
pathological features  of (\ref{lordireq}), like microscopic non-causality,
runaway solutions, preacceleration, and other bizarre effects$^{(4)}$. The
adoption of an integral equation with a convenient choice of limits can avoid
either one of these two last problems, but not both. On the other hand the
presence of this term  is necessary for the  maintenance of energy-momentum
conservation; without it it would be required a null radiance for an
accelerated charge. The argument, although correct, that such causality
violations are not observable because they are outside the scope of classical
physics$^{(5)}$ and are blurred$^{(6)}$  by quantum-mechanics effects is not
enough compelling,  because these same problems  remain in a quantum formalism,
just disguised in other apparently distinct problems.\\  It must be added that
the inclusion of spin and some extension  or structure for  the electron would
be just a complication without a changing in the  essence  of  the problem. Taking a spinless particle is
a valid simplifying hypothesis since the point at stake is not that one must
consider every property of the physical electron, but why one gets physically
non-acceptable results if  one  starts from apparently good premises  and  uses  only  mathematically sound procedures? It can only
mean that something in the premises or in
the procedures is not as good as one thinks.
The problems that appear in both classical and quantum theories when  one
has to consider the limit situation of two objects,  the  electron
and its electromagnetic field for example, defined in a same  point are the
crux of the question.  For  the classical electron the picture  is quite clear:
the energy-momentum
conservation produces  sound physical results in any region
around the charge except at the position of the charge.  This, it will be
argued in the following, is  an strong indication of the breaking down of the
validity  of  some accepted premises about the structure of the space-time:
electron and photon require different local space-time structure. The failure
of recognizing this results in equation (\ref{lordireq}). It amounts to
requiring that the propagation of a massive object (the electron) attend the
same constraint of the photon, a massless object.\\ The geometrization of a
physical principle is a very useful tool because it assures its automatic
implementation and allows that we concentrate our attention on other aspects of
the problem we are studying.  The Minkowski spacetime represents a
geometrization of the relativistic requirement that the velocity of light be a
universal constant. Despite its undisputed success through the Theory of
Special Relativity, in the interface  between a field and its source, it
produces a causality violating description. Revisiting the Lienard-Wiechert solution we show that its implicit causality
implementation can be also geometrized; it implies a structure of spacetime
more complex than the Minkowskian one and its light-cone structure. This
suggested new model of spacetime requires a revision of our concepts of field
theories of interacting massive and massless fields, a discussion to be started
in a subsequent paper, and shows the weak points in the demonstrations of the
Lorentz-
Dirac equation, which resumes our goals here.
\section{THE LORENTZ-DIRAC EQUATION}
 The derivation of the  Lorentz-Dirac  equation, with the  use  of
techniques of distribution theory, can be roughly schematized$^{(1)}$ in the
following way.
The electromagnetic field $F_{\mu\nu}=A_{\mu,\nu}-A_{\nu,\mu}$,  with
$\partial.A\equiv\partial_{\mu}A^{\mu}=0$, satisfies the
Maxwell's equations,  $\Box F = 4\pi J$,where J, given by,
\begin{equation}
J(x)=e\int d\tau V\delta^{4}[x-z(\tau)],
\end{equation}
is the  current  for  a  point
particle with electric charge e and  four-velocity  V.  
The Lienard - Wiechert solution$^{(3,5,7)}$,
\begin{equation}
\label{A}
A=\frac{eV}{\rho} ,\;\;\;\;\;\rho>0,
\end{equation}
in terms of retarded \nolinebreak coordinates, by which any spacetime point x
is constrained  with  a particle world-line point $z(\tau)$ by
\begin{equation}
\label{const}
R^{2}=0
\end{equation}
\nolinebreak and $R^{0}>0$, with $R\equiv x-z(\tau), \;  \;\;\rho\equiv
-V.\eta.R$, where $\eta$ is the  Minkowski metric tensor, (with signature +2).
$\rho$ is the spatial distance between the point  x where
the electromagnetic field is observed and the point $z(\tau)$, position
of the charge, in  the
charge rest frame at its retarded time.
The total particle and field energy momentum tensor, $T=T_{m}+\Theta$, consists
of $$T_{m}=m\int d\tau VV\delta^{4}[x-z(\tau)]$$
$$\Theta^{\mu\nu}=\frac{1}{4\pi}(F^{\mu\alpha}F_{\alpha}^{\:\nu}- \frac{\eta
^{\mu\nu}}{4}F^{\alpha\beta}F_{\beta\alpha}),$$  where $F=F_{ret}+F_{ext}$ is
the  retarded  field  added  to  any external electromagnetic field acting on
the charge. It induces $\Theta=\Theta_{ret}+\Theta_{mix}+\Theta_{ext}$. Mention
to some messy calculations$^{1}$ related to the highly non integrable parts of
$\Theta_{ret},$ which requires some renormalizations of $\Theta$ on the charge
worldline, are being omitted here.\\  The required momentum conservation,
\begin{equation}
\label{conserv}
T^{\mu\nu},_{\nu}=0,
\end{equation}
is satisfied without any problem at any point except at $\rho=0,\;\; (x=z)$,
where T is not defined.  In order to handle the singularity at $ \rho=0,$
T must be treated not as just a function defined only at $\rho>0$ but as a
distribution defined everywhere. Then,  (\ref{conserv})  is replaced  by
\begin{equation}
\label{leidecons}
\int dx^{4}T^{\mu\nu},_{\nu}\phi(x)=-\int
dx^{4}T^{\mu\nu}\phi,_{\nu}=-\lim_{\varepsilon\to0}{\int
dx^{4}T^{\mu\nu}\phi,_{\nu}\theta(\rho-\varepsilon})=0,
\end{equation}
where $\phi(x)$ is an arbitrary differentiable function with a compact support
and $\theta(x)$ is the Heavyside function, $\theta(x>0)=1$, $\theta(x<0)=0$.
Another integration by parts gives
\begin{equation}
\label{Tlimit}
\lim_{\varepsilon\to0}{\int
dx^{4}\rho,_{\nu}T^{\mu\nu}\phi(x)\delta(\rho-\varepsilon)}=0,
\end{equation}
which, after integration, produces, in the limit, the Lorentz-Dirac equation
(\ref{lordireq}). We want to pin point a crucial passage (common to most
derivation of this kind) in this procedure for posterior careful analysis: the
limit $\varepsilon\to0$, which represents a change from a point x where there
is only electromagnetic field and no electric charge, $\rho>0$, to a point
$z(\tau)$, instantaneous location of the electron, $\rho=0$.\\

\section{GEOMETRY OF CAUSALITY}
\noindent There is a beautiful and physically meaningful underlying geometry
describing the structure of causality in relativistic classical
electrodynamics, of which we will give here just a brief description.
The Lienard-Wiechert solution (\ref{A}) is an explicit function  of x and of
$\tau$, the retarded proper time of its source, solution of the constraint
(\ref{const}). When  taking  derivatives  of
functions of retarded  coordinates, like (\ref{A}), the differentiation of the
constraint (\ref{const}) must
be considered and it implies$^{(7,9)}$ on $R.dR=0$, or
\begin{equation}
\label{lconst}
d\tau+K.dx=0
\end{equation}
where $K\equiv R/\rho=-\partial \tau /\partial x$.  The  effects  of  this
constraints  on
derivatives of functions of retarded time, like A, can be automatically
accounted for if each derivative is replaced by a directional derivative,
\begin{equation}
\label{convective}
\partial _{\mu} \longrightarrow \nabla_{\mu}\equiv
\partial_{\mu}-K_{\mu}\partial/\partial\tau
\end{equation}
The constraints (\ref{const}) and (\ref{lconst}) have a clear geometrical and
physical meaning: the electron and its electromagnetic field must belong to and
remain in a same lightcone; they represent, respectively, a global and a local
implementation of the relativistic causality.
\noindent In the standard formalism, which we are reviewing, there is a clear
distinction between the treatment given to the electron and the one given to
its electromagnetic field. It is now convenient to adopt a notation where these
distinctions are reduced to the essentially necessary. So, we  change the
notation, replacing R by  $\Delta x$ and extending its meaning to be a change
in the location of a physical object (particles, fields, etc). Therefore, the
constraint $R^{2}=0$ is replaced by
\begin{equation}
\label{nulot}
\Delta x.\eta.\Delta x=0,
\end{equation}
showing, in an explicit way, that the electromagnetic field, as a massless
field, propagates keeping constant its propertime, $\Delta\tau=0$. K as a null
vector, $K^{2}=0$, represents a lightcone generator, the direction of
propagation of the electromagnetic field. Equation (\ref{const}) or
(\ref{nulot}) can be seen as a restriction  on the set of solutions of
(\ref{lconst}), and both (\ref{nulot}) and (\ref{lconst}), are constraints to
be imposed on the propagation of massless objects.  They make sense for the
electromagnetic field and as such they have accordingly been used in section 2
for $\rho >0$, but they cannot be extended to the propagation of a massive
object, like an electron.
\noindent The appropriate constraint, equivalent to (\ref{nulot}) for an
electron, has to be
\begin{equation}
\label{econst}
-(\Delta\tau)^{2}=\Delta x.\eta.\Delta x,
\end{equation}
where $\Delta\tau$ is the variation of the electron propertime during its
propagation along a distance $\Delta x$; likewisely the constraint
(\ref{lconst}) must be replaced by
\begin{equation}
\label{elconst}
d\tau+V.dx=0;
\end{equation}
and, similarly, a directional derivative corresponding to (\ref{convective}) is
defined, replacing K by V:
\begin{equation}
\label{econvective}
\partial _{\mu} \longrightarrow \nabla_{\mu}\equiv
\partial_{\mu}-V_{\mu}\partial/\partial\tau
\end{equation}
The differences between (\ref{convective}) and (\ref{econvective}) just
reflects the distinct constraints on the propagation of massive and of massless
physical objects.\\
We are now in condition to define the unifying geometric background that
underlies equations (\ref{convective}-\ref{econvective}). Consider all the
physical objects (electrons, electromagnetic fields, etc) immersed in a  flat
5-dimensional space, $R_{5}\equiv R_{4}\otimes R_{1}$, whose line elements are
defined by
\begin{equation}
\label{efive}
(\Delta S_{5})^{2}=\Delta x^{M}\eta_{\mbox{\tiny{MN}}}\Delta x^{N}= (\Delta
S_{4})^{2}-(\Delta x^{5})^{2}=\Delta x.\eta.\Delta x-(\Delta x^{5})^{2},
\end{equation}
where $M,N=1\;to\;5$. Immersed in this larger space, every physical object is
restricted to a 4-dimensional submanifold, its SPACETIME, by
\begin{equation}
\label{gconst}
-(\Delta x^{5})^{2}=\Delta x.\eta.\Delta x.
\end{equation}
$(\Delta S_{5})^{2}=-2(\Delta\tau)^{2}$ for a physical object, always. In other
words, the range of $x^{5}$ of a physical object is restricted to the range of
its very propertime, $\Delta x^{5}=\Delta\tau$. This is a causality condition,
standing for both (\ref{nulot}) and (\ref{econst}). So, the constraints on the
propagation of physical objects become restrictions on their allowed domain in
$R_{5}$, that is in the definition of their allowed spacetime. (\ref{gconst})
may be written, in an obvious notation, as
\begin{equation}
\label{hypercone}
(\Delta t)^{2}=(\Delta\tau)^{2}+(\Delta{\vec{x})}^{2},
\end{equation}
\noindent which defines a 4-dimensional hypercone in the local tangent space of
$R_{5}$. See figure 1. It is a CAUSALITY-CONE, a generalization of the Minkowski lightcone.

\begin{figure}
\vbox{{\centerline{\epsfsize=88cm\epsfbox{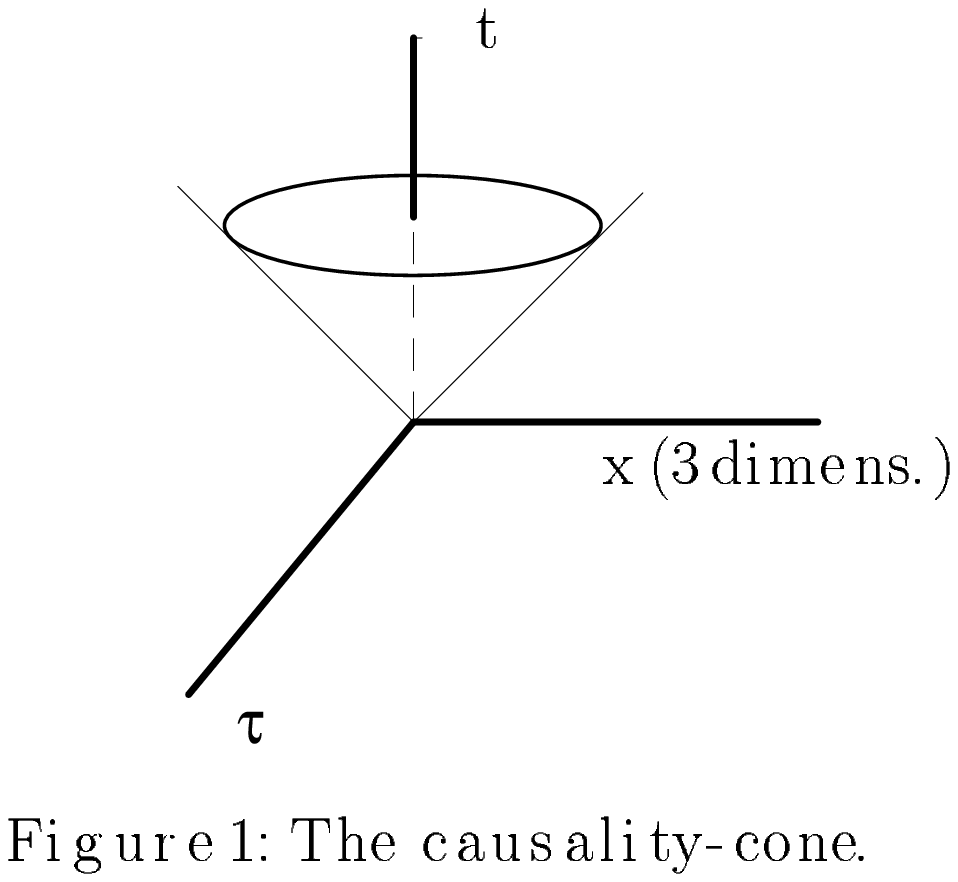}}}}
\caption{}
\end{figure}

 A lightcone, the domain of a massless physical object, is an intersection of a
causality-cone and a 4-dimensional hyperplane of RELATIVISTIC ABSOLUTE
SIMULTANEITY, defined by: $x^{5}=const$.  The interior of a lightcone is the
projection of a causality-cone on such a $(x^{5}=const)$-hyperplane. Each
observer perceives an strictly $(1+3)$-dimensional world and his $\Delta x^{5}$
coincides with the elapsed time measured on his own clock, as required by
special relativity; it represents his aging, according to his own clock.\\ This
is in contradistinction to Kaluza-Klein type of theory for unification of
fields, which uses a spacelike fifth dimension and then needs a
compactification mechanism to justify the non observability of $x^{5}$. The use
of a timelike fifth coordinate is, of course, not new in physics. See for
example the references [8,9] and the references therein. \\
A subtle detail must be observed. It is not correct that we are interpreting
$x^{5}$ as a proper time; it is $\Delta x^{5}$, the variation of $x^{5}$ of a
physical object, that is interpreted as the variation of {\it its} proper time,
its aging.
The propagation of physical objects, in this geometric setting, is restricted
by the differential of (\ref{gconst}), $\Delta\tau d\tau+\Delta x.dx=0$, or
\begin{equation}
\label{generator}
d\tau +f.dx=0,
\end{equation}
$f.dx=f_{\mu}dx^{\mu}$, where $f=\frac{\Delta x}{\Delta \tau},$ and f is a
timelike $4$-vector if $d\tau \ne 0$,
or (extending (\ref{generator}) to include) a light-like $4$-vector if $d\tau =
0$.  (\ref{generator}) defines a family of 4-dimensional hyperplanes
parameterized by f and enveloped by the causality-cone (\ref{hypercone}).
(\ref{hypercone}) and (\ref{generator}) define a causality-cone generator whose
tangent, projected on a $(\Delta x^{5}\!=\!constant)\!-\!hyperplane$, is f.
A lightlike f corresponds  to K of (\ref{lconst}) while a timelike f stands for
V of (\ref{elconst}).\\
Let us consider the figure 2 in order to have a clear understanding of the
meaning of $x^5$ of a physical object as its aging.

\begin{figure}
\vbox{{\centerline{\epsfsize=88cm\epsfbox{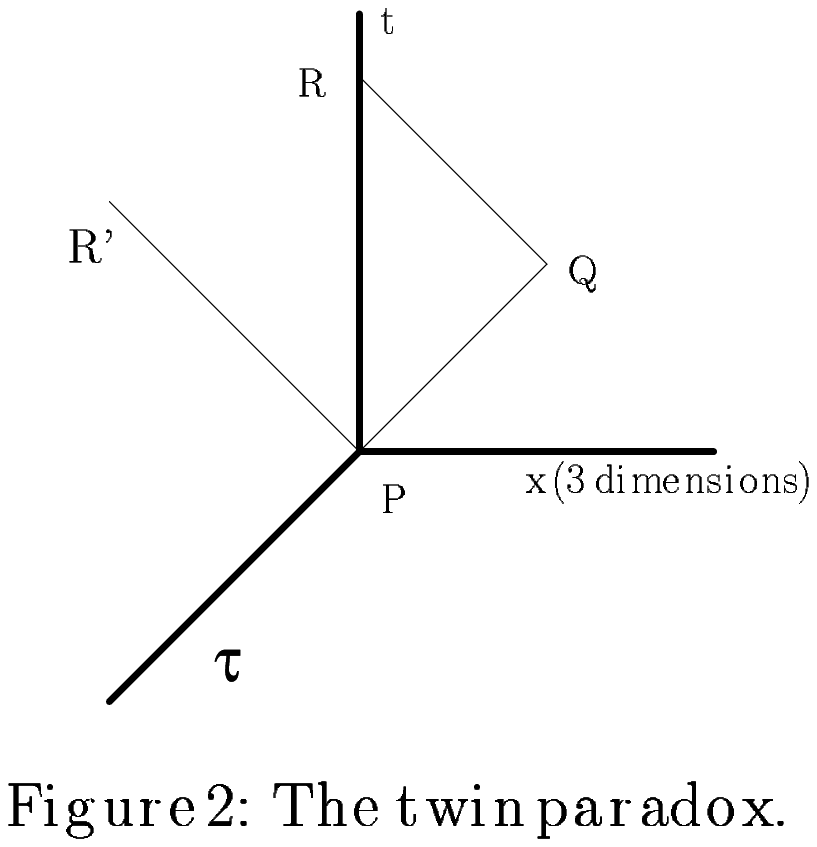}}}}
\caption{}
\end{figure}

This figure may represent a
vain physicist looking himself at a mirror, or the limiting case ($v\approx c$)
of the twin paradox in Special Relativity. $PR^{\prime }$ and PQ belong to a
same causality-cone. PQ belongs to the light-cone (taking $v\approx c)$. $PR'=(0,0,0,\Delta\tau,\Delta x^{5})$ with $\Delta t=\Delta \tau$, while $PQ=(\Delta{\vec{r}},\frac{\Delta t}{2},0)$ and $QR=(-\Delta{\vec{r}},\frac{\Delta t}{2},0).$
$PR^{\prime }$ is the physicist world-line on his rest frame. R is the
physicist image reflected (back to him) at Q, or his twin brother returning
from a trip to Q. They meet again at the time $t=t_{R}>0$, at the same space
point ($\vec{r}=0$, from where they had departed from each other,  but now with
distinct fifth coordinates, $x^{5}_{R}=0\;$ and $x^{5}_{R'}>0$, that represent
their distinct agings.\\ 
Let us mention now a rich and interesting point of this geometry. Observe the
difference between $\tau$ and t in (\ref{hypercone}): they are invariant under
different subgroups of isometry  ---$SO(3,1)$ and $O_{4}$, respectively--- of
the causality-cone.  Both sides of $$(\Delta \tau)^{2}=(\Delta t)^{2}+(\Delta {\vec{x}})^{2}$$
are invariant under transformation of the $S0(1,3)$ group, that is, rotation in a Minkowski spacetime $({\vec{x}},t)$; but in $$(\Delta t)^{2}=(\Delta \tau)^{2}+(\Delta {\vec{x}})^{2},$$ both sides are invariant under $0_{4}$, the rotation group in Euclidian 4-dimen{\-}sional spacetime $({\vec{x}},\tau).$\\ The use of t in the place of $\tau$ as the invariant
corresponds to Wick rotation without the need of analytic
continuation$^{(10,11)}$,  $t\rightarrow it$, and lends to it a clear physical and geometrical interpretation. 
  Physically it means that, for an Euclidian 4-dimensional spacetime, events should be labelled not by the time measured in the observer's clock, but with their local proper time, read on their local clocks.  $0_{4}$ is the invariance group of the causality-cone for rotations around its t-axis. 
Care must be taken with the interpretation of the $0_{4}$ sub-groups involving $\tau$, as they correspond to Lorentz and conformal transformations. \\
For those unaccustomed to the idea of extra-dimensions, we remind again that
the Minkowski spacetime represents the geometrization of an experimentally
founded physical principle: the constancy of the speed of light. It requires
that the time (up to then, just a parameter) be treated as the fourth
coordinate of a 4-dimensional manifold, the spacetime. We are doing here
something very similar: the geometrization of causality, embodied in the
relations (\ref{gconst}) or (\ref{hypercone}). It  requires  a fifth
coordinate with the role of a propertime.\\
 We can now return to our initial problem, which in this  geometry is pictured
by an electron and its electromagnetic field in a same causality-cone, each
along its own cone-generator. Up to here this is just another picturization of
this problem without any real change in its usual dynamics. The first real and
fundamental change appears when one considers the metric structure for the
spacetime of each physical object. The metric induced by (\ref{generator}) on
the  spacetime  of a physical object, its 4-dimensional submanifold,
$(dS_{5})^{2}=dx.\eta.dx-(f.dx)^{2}=dx.(\eta-ff).dx$, is given by
$g_{\mu\nu}=\eta_{\mu\nu}$ for a massless field (since then $d\tau=-f.dx=0$),
and by $g_{\mu\nu}=\eta_{\mu\nu}-f_{\mu}f_{\nu}$, and
$g^{\mu\nu}=\eta^{\mu\nu}+\frac{f^{\mu}f^{\nu}}{2}$ for massive physical
objects.
The distinct causality requirements of  massive  and  of  massless
fields and particles are,  therefore,  represented  by  immersions
with distinct metric structures. They can both be written in a single
expression (using either $f^{2}=0$ or $f^{2}=-1$ for, respectively massless and
massive objects):
\begin{eqnarray}
g_{\mu\nu}&=&\eta_{\mu\nu}+f^{2}f_{\mu}f_{\nu},\\
\label{metric}
g^{\mu\nu}&=&\eta^{\mu\nu}-f^{2}\frac{f^{\mu}f^{\nu}}{1+(f^{2})^{2}}.
\end{eqnarray}
At this point we can understand that a plus sign in front of $(\Delta\tau)^{2}$
of the  line element (\ref{efive}) would imply on $\Delta{\cal S}_{5}\equiv0$
for all physical object and would induce
$g_{\mu\nu}=\eta_{\mu\nu}+f_{\mu}f_{\nu}$, as a metric on a causality-cone of a
massive object, which would not be consistent because then,
$g_{\mu\nu}f^{\mu}\equiv0$.\\
The existence of two distinct metric structures for a massive and a massless
field invalidates (\ref{lordireq}) as the result of (\ref{leidecons}).
While $T^{\mu\nu},_{\nu}=0$ for $\rho>0$ remains valid in this new picture, its
limit when $\rho\rightarrow 0$ is not as simple as described before because  it
 involves  now  a  local change of manifolds with different metric structure
($\eta\rightarrow\eta-ff$).
Physically it only makes  sense!  For $\rho>$0  one is dealing  with
electromagnetic fields (photons)  for  which (\ref{lconst},\ref{nulot})
represent the causality requirement that A(x,t) and $z(\tau_{\mbox{\tiny
ret}})$ remain on a same
light-cone, but for $\rho=0$ one has an electron, a massive particle,
which must attend a completely different causality relation
(\ref{econst},\ref{elconst}). As a matter of fact, the limit  of  K
when $\rho$ tends to zero is an indeterminacy of the type 0/0 that can
be resolved  with a  derivative $d/d\tau$ and the L'Hospital rule:
\begin{equation}
\label{limit}
\lim_{\rho\to0}{K_{\mu}}=\lim_{\rho\to0}{f_{\mu}}=V^{\nu}\eta_{\nu\mu}
\equiv\stackrel{~}{V_{\mu}}.
\end{equation}
This is coherent with (\ref{convective}) and (\ref{econvective}).
The geometric meaning of (\ref{limit}) can easily be understood if we remind
that K, as a 4-vector tangent to the lightcone, can be written as $K=V+N$, with
$V.N=0$.
This change of K to V was not considered  in the limit passage of
(\ref{Tlimit}), as also, of course, the change of metric required by
(\ref{metric}).  This procedure extends, in fact, the photon causality
constraint (\ref{nulot}) to the electron; it corresponds to treating the
electron as if it were a massless object. This new vision of spacetime requires
a revision not only of the Lorentz-Dirac equation but of any theory of
interacting fields. This will be done in a subsequent paper. Our immediate goal
has been attained with the stressing of the connections among causality
violation in the Lorentz-Dirac equation and the spacetime structure.
\section{SUMMARY AND CONCLUSIONS}
The Lienard-Wiechert solutions are closely related to the Lorentz-Dirac
equation, but while the first have a well drawn picture of causality
preservation, based on the light-cone structure of the Minkowski spacetime, the
second is, nonetheless, well known for its problematic causality violating
solutions. For this reason, this equation has always been accompanied by many
doubts about its validity. But it has been obtained from the most diverse
approaches and its uniqueness has been scrutinized and proved under very
generic and acceptable conditions$^{(1,2)}$. However, we do not endorse the
apparently most accepted view that this is, after all, the correct equation and
that its problems appear only when we stretch its application to situations
when a quantum theory should be used instead.\\
With the strategy of geometrizing the Principle of Causality, that is, of
transferring its implementation to the background spacetime structure, we find
that a model of spacetime, more complex than the Minkowski's one, is required.
It makes clear that the weak point common to all demonstrations of the
Lorentz-Dirac equation is the extrapolation for the electron of restrictions
that are valid only for its electromagnetic field. The Minkowski spacetime
represents just a geometrization of the Einstein postulates of Special
Relativity, and so, it does not contemplate the difference in the metric
structure required for a geometric implementation of causality. Therefore, the
Lorentz-Dirac equation is the result of imposing to the electron a causality
behaviour that is valid only for its electromagnetic field. Our next step is to
show that if these distinct metric structures are taken into consideration, the
Schott term does not appear in the equation of motion of a classical point!
 electron without any jeopardy to
energy conservation. But this belongs to a subsequent paper.

\end{document}